\documentclass[aps,prb,10pt,twocolumn,longbibliography,superscriptaddress,amsmath,amssymb,floatfix]{revtex4-2}

\usepackage[utf8]{inputenc}
\usepackage{graphicx}
\usepackage{bm}
\usepackage{mathrsfs}
\usepackage{dcolumn}
\usepackage{makecell}
\usepackage{booktabs}
\usepackage[colorlinks,linkcolor=blue,citecolor=blue,urlcolor=blue]{hyperref}
\usepackage{xcolor}

\newcommand{\T}{\mathcal{T}}

\newcommand{\Ez}{E_z}

\newcommand{\LTEcell}[2]{\parbox[t]{#1}{\raggedright #2}}

\begin{document}

\title{Electrically Reversible Layer-Tensor Responses in Bilayers}

\author{Zehou Li}
\affiliation{Key Laboratory of Micro-nano Energy Materials and Application Technologies,
  University of Hunan Province $\&$ College of Physics and Electronic Engineering,
  Hengyang Normal University, Hengyang 421002, China}
\author{Shenda He}
\affiliation{School of Science, Hunan Institute of Technology, Hengyang 421002, China}
\author{Pan Zhou}
\email{zhoupan71234@xtu.edu.cn}
\affiliation{School of Materials Science and Engineering, Xiangtan University,
  Xiangtan 411105, China}
\author{Baoru Pan}
\affiliation{School of Materials Science and Engineering, Xiangtan University,
  Xiangtan 411105, China}
\author{Rui Tan}
\email{rtan@hynu.edu.cn}
\affiliation{Key Laboratory of Micro-nano Energy Materials and Application Technologies,
  University of Hunan Province $\&$ College of Physics and Electronic Engineering,
  Hengyang Normal University, Hengyang 421002, China}
\author{Lizhong Sun}
\email{lzsun@xtu.edu.cn}
\affiliation{School of Materials Science and Engineering, Xiangtan University,
  Xiangtan 411105, China}

\date{September 19, 2026}

\begin{abstract}
The layer Hall effect provides a prototypical example of a hidden layer response in van der Waals bilayers: opposite layers carry compensating Hall currents at zero out-of-plane electric field, while an applied out-of-plane electric field $E_z$ selects one layer over the other and switches the net Hall response. Here we show that this field-switching mechanism is not restricted to Hall transport, but follows from a general symmetry principle for response tensors. We formulate a selection rule for the layer tensor effect (LTE) using Jahn symbols. For a tensor with Jahn symbol $J$ and rank $r$, the stacking operation that exchanges the two layers acts as a coset representative relating the $+E_z$ and $-E_z$ states. Tensor components that are odd under this operation are forced to cancel between layers at $E_z=0$, and acquire a macroscopic response satisfying $T_\alpha(+E_z)=-T_\alpha(-E_z)$. This criterion allows us to enumerate the symmetry-allowed electrically reversible tensor components associated with different Jahn symbols and to identify their physical realizations in common response tensors. We apply the theory to representative anomalous Hall, shift-current, nonlinear Drude, piezoelectric, spin Hall, and spin-photocurrent channels. Model calculations and first-principles simulations further confirm the predicted field-odd responses. Our results establish an electrically switchable LTE framework, providing a unified route to multifunctional layertronic responses in bilayer materials.
\end{abstract}

\maketitle

\textit{Introduction}---Physical responses in crystalline solids are described by tensors mapping external perturbations to induced observables, e.g., linear conductivity $j_i=\sigma_{ij}E_j$, nonlinear conductivity $j_i=\chi_{ijk}E_jE_k$, and piezoelectric response $P_i=d_{ijk}u_{jk}$. By Neumann’s and Curie’s principles, these tensors must be invariant under the crystallographic or magnetic point group, which strictly constrains their allowed, forbidden, and dependent components~\cite{Nye1957}. This point-group constraint holds independent of microscopic origin, and any Kubo-type formula automatically satisfies it~\cite{Xiao2010,Nagaosa2010,Zhang2023Dictionary}. The antiunitary (time-reversal) part is more subtle: for nonequilibrium transport and optical responses it must be interpreted through the appropriate Onsager--Casimir relation rather than assigned a fixed scalar parity~\cite{Gallego2019}. For systems endowed with additional discrete internal degrees of freedom, symmetry operations act simultaneously on real-space coordinates and internal labels, giving rise to extra symmetry constraints and enabling response phenomena absent in conventional bulk crystals~\cite{layer5,layer2,layer1,Feng2023SpinValleyLayerAHE}. 

The layer index is a natural internal degree of freedom in van der Waals bilayers. 
In $\mathcal{PT}$-symmetric but time-reversal-breaking bilayers, such as antiferromagnetic MnBi$_2$Te$_4$, symmetry forbids a net Berry curvature, while allowing finite and opposite layer-resolved Berry curvatures. 
These hidden Berry-curvature textures~\cite{Zhang2014HiddenSpin,Riley2014HiddenSpin,Cho2018HiddenBerry,Arnoldi2024HiddenSpin} give rise to the layer Hall effect: at $E_z=0$, anomalous Hall currents from opposite layers cancel, whereas an out-of-plane electric field breaks the layer balance and produces a net Hall response that switches sign under $E_z\to -E_z$~\cite{Xiao2012Valley,Gao2021LayerHall,Chen2024LayerHall,Feng2023LayerHall,Zhang2023LayerAHE,Li2024Layertronics,Liu2024QALHE,Tian2025QALHE}. 
The electrical tunability of layer-resolved responses~\cite{Yasuda2021,ViznerStern2021,Wang2022FerroelectricTMD,Wu2021SlidingFE} then raises a broader question: can the same hidden-layer mechanism be promoted from anomalous Hall transport to general tensor responses? 
If so, van der Waals bilayers would offer a generic platform for electrically reversible control of diverse physical responses.

Here we develop a general symmetry framework demonstrating that electrically reversible hidden-layer physics generalizes to a broad class of layer-resolved tensor responses, which we term the layer tensor effect (LTE). We derive a layer-tensor selection rule for bilayers in which the $+E_z$ and $-E_z$ states are connected by a layer-exchanging stacking operation $\hat{\ell}$. For a tensor component $T_\alpha$, specified by a Jahn symbol $J$, the response is characterized by a component-dependent eigenvalue $\lambda_\alpha$.  Components with $\lambda_\alpha=-1$ are layer-odd: they cancel between the two layers at $E_z=0$, become finite once the layer-exchange symmetry is broken, and are symmetry-enforced to switch sign under field reversal, $T_\alpha(+E_z)=-T_\alpha(-E_z)$. Concrete symmetry analyses show that such layer-dependent and electrically reversible behavior can occur in shift current~\cite{Young2012Shift,Tan2016BPVE,lte_ahn_prx_2020,lte_zhao_disorder_shift_2025,Qian2023Shift2DFE,Yang2024GiantShift,Li2021CuInP2S6BPVE,Zhang2019MagneticBPVE,Cheng2021SeTeBPVE}, nonlinear Drude~\cite{Wang2024Ohmic,NagaosaYanase2024,lte_liu_switchable_2025}, nonlinear Hall~\cite{Sodemann2015,lte_matsyshyn_sodemann_2019,Ma2019NLH,Kang2019,Du2021NLH,NLHE10,QMD3}, piezoelectric~\cite{Duerloo2012,Wu2014Piezo}, spin Hall~\cite{Sinova2015}, and spin-photocurrent responses~\cite{Young2013SpinBPVE,Fang2024NeelSpin,Xiao2021SpinPhotogalvanic}. Model calculations and first-principles simulations further confirm these predictions, establishing a general route toward electrically switchable layertronic tensor responses.

\textit{Layer-exchange symmetry and electrically switchable tensors}---We begin by specifying the operation that relates the two layers.  A bilayer is
constructed from a reference layer $L$ and a partner layer
$L'=\hat{\ell}L$, where the layer-exchange stacking operation is written in
Seitz form as
\begin{equation}
    \hat{\ell}=\{R_\ell|\mathbf{t}_\ell\}\mathcal{T}^{\theta_\ell}.
\label{eq:seitz-stack}
\end{equation}
Here, $R_\ell$ is the point operation, $\mathbf{t}_\ell$ is the stacking
translation, $\mathcal{T}$ is time reversal, and
$\theta_\ell=0,1$ specifies whether time reversal is included. Typical choices for $R_\ell$ in bilayers are inversion, an in-plane twofold rotation, and a mirror plane normal to the $z$ axis. The zero-field parent bilayer is $B_0=L+\hat{\ell}L$.
If an out-of-plane electric field is applied, the layer symmetry $\hat{\ell}$ can be broken by inducing an electrostatic potential difference between the two constituent layers. Here we concentrate on the two oppositely biased states $B(+E_z)$ (upward field) and $B(-E_z)$ (downward field) that remain connected via a subset of the layer-exchange symmetry operation.

\begin{figure}[!tbp]
\includegraphics[width=\columnwidth]{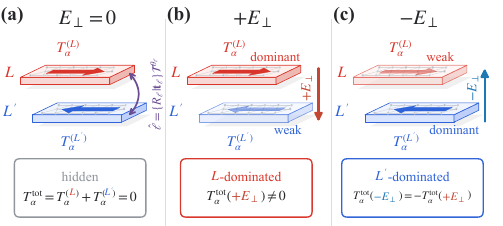}
\caption{\label{fig:lte-schematic}
Electrically switchable LTE for a component $T_\alpha$ that is
layer odd under the layer-exchange stacking operation
$\{R_\ell|\mathbf{t}_\ell\}$, so the two layers carry equal and opposite
contributions. The perpendicular field is denoted by $E_\perp=E_z$ in the schematic. (a)~At $E_z=0$ they cancel, $T_\alpha^{\mathrm{tot}}=0$
(hidden).  (b)~$+E_z$ makes one layer dominant and the other weak, giving
$T_\alpha^{\mathrm{tot}}(+E_z)\neq0$.  (c)~$-E_z$ swaps the dominant and
weak layers, reversing the sign,
$T_\alpha^{\mathrm{tot}}(-E_z)=-T_\alpha^{\mathrm{tot}}(+E_z)$.}
\end{figure}

Figure~\ref{fig:lte-schematic} illustrates the resulting layertronic mechanism
for a general tensor response.  We denote a component of a rank-$r$ response
tensor by the multi-index $\alpha=(i_0i_1\cdots i_{r-1})$, so that $T_\alpha\equiv T_{i_0i_1\cdots i_{r-1}}$.  Suppose that the constituent layer is allowed by symmetry to carry a finite local component $T_\alpha^{(L)}$.  The stacking operation generates the corresponding component in the partner layer, $T^{(L')}_\alpha=\bigl[\mathcal{D}_J(\hat{\ell})T^{(L)}\bigr]_\alpha$, where $\mathcal{D}_J$ is the tensor representation determined by the stacking operation.  For a layer-odd component, the two local responses are equal in
magnitude and opposite in sign. Hence, at $E_z=0$,
\begin{equation}
    T_{\alpha}^{\mathrm{tot}}(0)
    =
    T_{\alpha}^{(L)}+T_{\alpha}^{(L')}
    =
    0 ,
\label{eq:hidden-layer-tensor}
\end{equation}
although the layer-resolved tensors themselves remain finite.  This is the hidden layer-tensor state.

As shown in Figs.~\ref{fig:lte-schematic}(b) and~\ref{fig:lte-schematic}(c), an out-of-plane electric field can couple to the layer degree of freedom, thereby breaking the layer-exchange symmetry that relates $L$ and $L'$. The field does not create a new tensor channel. Instead, it selects one of two oppositely locked
layer contributions.  For $+E_z$, the low-energy electronic states of one
layer can dominate the response, giving a finite macroscopic tensor
$T_\alpha^{\mathrm{tot}}(+E_z)$.  Reversing the field exchanges the
dominant layer, so that a layer-odd component satisfies
\begin{equation}
    T_{\alpha}^{\mathrm{tot}}(+E_z)
    =
    -
    T_{\alpha}^{\mathrm{tot}}(-E_z).
\label{eq:field-odd-response}
\end{equation}
This establishes the basic principle of electric-field activation and reversal
of hidden layer-resolved tensor responses in van der Waals bilayers, as shown in Fig.~\ref{fig:lte-schematic}.

\textit{Symmetry criterion}---
We now formulate this picture as a selection rule. For a tensor $T_\alpha$, its symmetry character is specified by its Jahn symbol $J$, which encodes its rank, polar or axial nature, index symmetry~\cite{Jahn1949,Gallego2019}. Consequently, the same crystal symmetry operation can impose distinct constraints on different physical tensors. For a response sector with definite Onsager parity, we set
$a_J=1$ when $J$ begins with $a$, and $a_J=0$ otherwise.  We also set $p_J=1$ when the spatial part of $J$ contains the axial constant $e$, and $p_J=0$ otherwise. The scalar time-reversal factor $(-1)^{a_J}$ is legitimate only when the response forms a closed Onsager sector of definite parity; whether this holds is fixed by the microscopic response kernel, not by the Jahn symbol, and is analyzed in Sec.~II of the SM~\cite{SM}.

For a magnetic space-group operation $g=\{R_g|\mathbf{t}_g\}\mathcal{T}^{\theta_g}$,
the homogeneous tensor transforms according to the point and time-reversal
parts of $g$,
\begin{equation}
\begin{split}
  [\mathcal{D}_J(g)T]_{i_0\cdots i_{r-1}}
  =&\,
  (-1)^{a_J\theta_g}
  (\det R_g)^{p_J}\\
  &\times R_{g,i_0j_0}\cdots R_{g,i_{r-2}j_{r-2}}
  R_{g,i_{r-1}j_{r-1}}\\
  &\times
  T_{j_0\cdots j_{r-1}},
\end{split}
\label{eq:tensor-rep}
\end{equation}
followed, when required by $J$, by the appropriate symmetrizer or
antisymmetrizer acting on equivalent indices.  The translational part
$\mathbf{t}_g$ does not change the local Cartesian tensor law, but it is
essential for determining which bilayer stacking symmetries survive.
For nonequilibrium kernels not closed under time reversal, the antiunitary action requires a response-specific Onsager map, as detailed in Sec.~II of the SM~\cite{SM}.

For an isolated monolayer with symmetry group $G_{\rm m}$, Neumann's
principle requires $T=\mathcal{D}_J(g)T$ for every $g\in G_{\rm m}$.
Equivalently, the allowed local tensor space is the image of the projector $\mathcal{P}_{G_{\rm m},J}=|G_{\rm m}|^{-1}\sum_{g\in G_{\rm m}}\mathcal{D}_J(g)$.
Thus the first requirement for an LTE channel is that the monolayer itself can support the component:
\begin{equation}
    \mathcal{P}_{G_{\rm m},J}\,u_\alpha^{J}\neq 0 ,
\label{eq:mono-allowed}
\end{equation}
where $u_\alpha^{J}$ is the unit tensor associated with the multi-index
$\alpha$, including the intrinsic symmetrization imposed by the Jahn symbol.
This condition removes trivially forbidden responses: a bilayer cannot hide a
tensor component that is already forbidden in each constituent layer.

The second requirement concerns how the same allowed tensor component
transforms under the layer-exchange stacking operation $\hat{\ell}$.  The local
tensor in the partner layer is obtained from that in the reference layer as $T^{(L')}=\mathcal{D}_J(\hat{\ell})T^{(L)}$.
The layer parity of a selected component is obtained by acting with
$\mathcal{D}_J(\hat{\ell})$ on its symmetrized unit tensor $u_\alpha^{J}$.
A component is an eigenchannel of
$\hat{\ell}$ when the full tensor representation returns it up to a scalar,
\begin{equation}
    \mathcal{D}_J(\hat{\ell})\,u_\alpha^{J}
    =
    \lambda_\alpha(\hat{\ell})\,u_\alpha^{J},
    \qquad
    T^{(L')}_\alpha
    =
    \lambda_\alpha(\hat{\ell})\,T^{(L)}_\alpha ,
\label{eq:eigenchannel}
\end{equation}
and the eigenvalue $\lambda_\alpha(\hat{\ell})=\pm1$ is the layer parity.
Equation~(\ref{eq:eigenchannel}) is the rigorous, frame-independent definition.
When the point part $R_\ell$ is diagonal in the chosen Cartesian
frame---as for $\mathcal{P}$, $C_{2x}$, $C_{2y}$, $C_{2z}$, $M_x$, $M_y$, and
$M_z$---every Cartesian component is automatically an eigenchannel, and
Eq.~(\ref{eq:eigenchannel}) reduces to the closed form
\begin{equation}
  \lambda_\alpha(\hat{\ell})
  =
  (-1)^{a_J\theta_\ell}
  (\det R_\ell)^{p_J}
  \prod_{m=0}^{r-1}(R_\ell)_{i_m i_m}.
\label{eq:lambda}
\end{equation}
When $R_\ell$ is non-diagonal in this frame---as for $C_{3z}$,
$C_{4z}$, $C_{6z}$, or $S_{4z}$---the diagonal product in
Eq.~(\ref{eq:lambda}) does not apply, even if the component happens not
to mix with others. The layer parity must then be read directly from the full
matrix $\mathcal{D}_J(\hat{\ell})$ through Eq.~(\ref{eq:eigenchannel}).  If, in
addition, $\hat{\ell}$ mixes several allowed components, one diagonalizes
$\mathcal{D}_J(\hat{\ell})$ within the symmetry-allowed tensor subspace, and the
$\lambda=-1$ eigenvectors define the layer-odd combinations.

To apply the criterion, one scans $R_\ell\mathcal{T}^{\theta_\ell}$ over the
candidate layer-exchange magnetic point operations and keeps those for which the
selected component is layer odd, $\lambda_\alpha(\hat{\ell})=-1$.  This
layer-parity test is applied only after the same component is confirmed to be
allowed in a fixed-$\Ez$ monolayer by Eq.~(\ref{eq:mono-allowed}).

The zero-field bilayer response of an eigenchannel is then
\begin{equation}
    T^{\rm bilayer}_\alpha(0)
    =
    \left[1+\lambda_\alpha(\hat{\ell})\right]
    T^{(L)}_\alpha .
\label{eq:zero-cancel}
\end{equation}
Therefore, $\lambda_\alpha(\hat{\ell})=-1$ is the defining criterion for a hidden layer tensor: the response is finite in
each layer but cancels in the unbiased bilayer.  Since the same layer-exchange
operation maps the two biased configurations into each other, the macroscopic
responses satisfy
\begin{equation}
    T_\alpha(+E_z)
    =
    \lambda_\alpha(\hat{\ell})T_\alpha(-E_z).
\label{eq:field-switch}
\end{equation}
Consequently, a layer-odd tensor is electrically switchable,
\begin{equation}
    T_\alpha(+E_z)
    =
    -
    T_\alpha(-E_z),
\label{eq:odd-field}
\end{equation}
whereas a layer-even tensor is unchanged by field reversal.  These relations show that the sign switching is determined by the layer-exchange character of the tensor component, independent of the microscopic formula used to compute the response.

\begin{table*}[t]
\caption{\label{tab:stacking}
Representative electrically switchable layer-tensor channels. The columns list a representative Cartesian component with its associated physical response, the Jahn symbol $J$ (encoding rank, polar or axial character, and index symmetry), the magnetic point operations that allow the component in a monolayer under a fixed $E_z$ [Eq.~(\ref{eq:mono-allowed})], and the layer-exchange operations that make the component layer-odd ($\lambda_\alpha=-1$), so that it remains hidden at $E_z=0$ but is switched on by an out-of-plane field. Entries generated by a unitary stacking operation are exact; antiunitary entries additionally require the response to form a closed Onsager sector of definite parity, a property fixed by its microscopic kernel rather than by the Jahn symbol, as detailed in Sec.~II of the SM~\cite{SM}.}
\footnotesize
\renewcommand{\arraystretch}{1.22}
\setlength{\tabcolsep}{2pt}
\begin{tabular}{@{}llll@{}}
\midrule\midrule
\LTEcell{0.22\textwidth}{Tensor component}
  & \LTEcell{0.11\textwidth}{Jahn symbol}
  & \LTEcell{0.35\textwidth}{Single-layer operations with $E_z$ }
  & \LTEcell{0.295\textwidth}{Bilayer stacking operations} \\
\midrule
\LTEcell{0.22\textwidth}{$\varepsilon^{xy}$, static dielectric}
  & \LTEcell{0.11\textwidth}{$[V^2]$}
  & \LTEcell{0.35\textwidth}{$C_{2z}$, $\T$, $C_{2z}\T$}
  & \LTEcell{0.295\textwidth}{$C_{2x}$, $C_{2y}$, $S_{4z}$,
    $C_{2x}\T$, $C_{2y}\T$, $S_{4z}\T$} \\
\LTEcell{0.22\textwidth}{$\sigma^{xy}$, anomalous Hall}
  & \LTEcell{0.11\textwidth}{$a\{V^2\}$}
  & \LTEcell{0.35\textwidth}{$C_{2z}$, $C_{3z}$, $C_{4z}$,
    $C_{6z}$, $M_x\T$, $M_y\T$}
  & \LTEcell{0.295\textwidth}{$C_{2x}$, $C_{2y}$, $\mathcal{P}\T$,
    $M_z\T$, $S_{4z}\T$} \\
\LTEcell{0.22\textwidth}{$d^{xyy}$, piezoelectric}
  & \LTEcell{0.11\textwidth}{$V[V^2]$}
  & \LTEcell{0.35\textwidth}{$C_{3z}$, $M_y$, $\T$, $M_y\T$}
  & \LTEcell{0.295\textwidth}{$\mathcal{P}$, $C_{2y}$, $\mathcal{P}\T$, $C_{2y}\T$} \\
\LTEcell{0.22\textwidth}{$\sigma^{yxx}$, shift current}
  & \LTEcell{0.11\textwidth}{$V[V^2]$}
  & \LTEcell{0.35\textwidth}{$C_{3z}$, $M_x$, $\T$, $M_x\T$}
  & \LTEcell{0.295\textwidth}{$\mathcal{P}$, $C_{2x}$, $\mathcal{P}\T$, $C_{2x}\T$} \\
\LTEcell{0.22\textwidth}{$\chi^{yxx}$, nonlinear Drude}
  & \LTEcell{0.11\textwidth}{$aV[V^2]$}
  & \LTEcell{0.35\textwidth}{$C_{3z}$, $M_x$, $C_{2z}\T$,
    $C_{6z}\T$, $M_y\T$}
  & \LTEcell{0.295\textwidth}{$\mathcal{P}$, $C_{2x}$, $C_{2y}\T$, $M_z\T$} \\
\LTEcell{0.22\textwidth}{$\sigma_s^{yxx}$, spin-photocurrent}
  & \LTEcell{0.11\textwidth}{$eV[V^2]$}
  & \LTEcell{0.35\textwidth}{$C_{3z}$, $M_y$, $\T$, $M_y\T$}
  & \LTEcell{0.295\textwidth}{$C_{2x}$, $M_z$, $C_{2x}\T$, $M_z\T$} \\
\LTEcell{0.22\textwidth}{$C^{xxxy}$, elastic stiffness}
  & \LTEcell{0.11\textwidth}{$[V^2][V^2]$}
  & \LTEcell{0.35\textwidth}{$C_{2z}$, $C_{4z}$, $\T$, $C_{2z}\T$,
    $C_{4z}\T$}
  & \LTEcell{0.295\textwidth}{$C_{2x}$, $C_{2y}$, $C_{2x}\T$, $C_{2y}\T$} \\
\midrule\midrule
\end{tabular}
\end{table*}

\textit{Enumeration of LTE channels}---Building on the symmetry-based requirements established above, we systematically classify tensor components that exhibit LTE across different bilayer stacking configurations. We refer to a locally allowed tensor component as an LTE channel when it is odd under the layer-exchange operation, so that the two layers carry equal and opposite responses in the symmetric bilayer. For a tensor defined by a given Jahn symbol, the analysis follows a two-step rule: we first identify the in-plane components that remain symmetry-allowed in a field-polarized monolayer, then determine whether each component acquires odd layer parity under a layer-exchange stacking operation. Since the dominant physical responses in condensed matter systems are captured by rank-2 through rank-4 tensors, we focus our systematic enumeration on this order range. The independent in-plane components are listed in Tables~S1--S3, and the complete LTE selection rules in Tables~S4--S6 of the Supplemental Material (SM)~\cite{SM}. Representative channels for each tensor rank are summarized in Table~\ref{tab:stacking}. Quantitatively, we find 28, 80, and 177 independent Jahn-resolved LTE channels for ranks 2, 3, and 4, respectively. This set of Jahn symbols covers the vast majority of linear and nonlinear response phenomena encountered in condensed matter systems.

We illustrate these selection rules and their physical consequences using several representative examples. We first consider the symmetric off-diagonal static dielectric component $\varepsilon^{xy}=\varepsilon^{yx}$, which describes the transverse polarization induced by a quasistatic external electric field. This component belongs to the Jahn-symbol class $[V^2]$. It is symmetry allowed in a monolayer invariant under $C_{2z}$, time reversal $\mathcal T$, or their combined operation, while it changes sign under layer exchange generated by a $C_{2x}$- or $C_{2y}$-type stacking operation. Consequently, the two layers carry opposite contributions to $\varepsilon^{xy}$ in the corresponding bilayer configurations. As a contrasting example, the antisymmetric Hall conductivity, $\sigma^{xy}=-\sigma^{yx}$, is characterized by the Jahn symbol $a\{V^2\}$. The time-reversal-odd prefix $a$ implies that this response requires broken time-reversal symmetry. Owing to its antisymmetric tensor character, $\sigma^{xy}$ remains compatible with an out-of-plane rotation $C_{nz}$ and with the antiunitary mirror symmetries $M_x\mathcal T$ or $M_y\mathcal T$. Its layer-resolved sign is reversed by stacking operations whose point parts are $C_{2x}$, $C_{2y}$, $\mathcal P\mathcal T$, $M_z\mathcal T$, or $S_{4z}\mathcal T$, thereby producing opposite Hall responses in the two layers.

We next use the polar rank-3 Jahn symbol $V[V^2]$ to show that both the symmetry allowance of a monolayer tensor and its parity under layer exchange are strongly related to the numbers of $x$ and $y$ indices. For any in-plane polar tensor component, an operation that reverses $x$ contributes a factor $(-1)^{N_x}$, whereas one that reverses $y$ contributes a factor $(-1)^{N_y}$, where $N_x$ and $N_y$ denote the numbers of $x$ and $y$ indices, respectively. When the operation is a symmetry of an isolated monolayer, only components with a total factor $+1$ are allowed. When the same operation exchanges the two layers, the factor instead determines the layer parity, with $-1$ indicating opposite tensor components in the two layers. Accordingly, among the six independent in-plane components of $V[V^2]$, a $C_{2x}$ layer exchange, which reverses $y$, assigns odd layer parity to the components containing an odd number of $y$ indices, namely $xxy$, $yxx$, and $yyy$, whereas a $C_{2y}$ exchange assigns odd parity to $xxx$, $xyy$, and $yxy$, which contain an odd number of $x$ indices. An inversion-type exchange reverses every in-plane index and therefore changes the sign of all polar rank-3 components because $N_x+N_y=3$ is odd. For example, the shift-current component $\sigma^{yxx}$, describing a current along $y$ driven by two $x$-polarized optical fields, has $N_x=2$ and $N_y=1$: it is invariant under $M_x$, for which the two $x$ indices produce a positive factor, but is reversed and thus forbidden by $M_y$, owing to its single $y$ index. Under a $C_{2x}$ layer exchange, the same odd number of $y$ indices gives $\lambda_{yxx}=(-1)^{N_y}=-1$, so the two layers carry opposite local shift-current tensors and their macroscopic responses cancel at zero out-of-plane field. The same counting rule applies at any tensor rank. For instance, the elastic stiffness component $C^{xxxy}\in[V^2][V^2]$ contains three $x$ indices and one $y$ index: it is invariant under $C_{2z}$, because all four indices are reversed and yield the factor $(-1)^{N_x+N_y}=+1$, but it is odd under a $C_{2x}$ layer exchange because $N_y=1$. Thus, the numbers of $x$ and $y$ indices provide a unified criterion for identifying both symmetry-allowed monolayer responses and hidden layer-odd channels in higher-rank tensors such as shift conductivity, elastic stiffness, and elastoresistivity.

\begin{figure}[!tbp]
\includegraphics[width=\columnwidth]{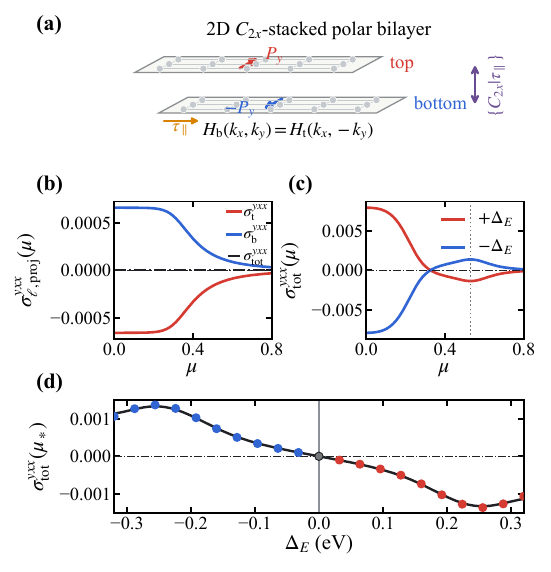}
\caption{\label{fig:2d-model}
Minimal tight-binding realization of a hidden, electrically switchable layer
shift current.  (a)~Two $C_{2x}$-related layers carrying opposite in-plane polar
distortions ($\pm P_y$), coupled by the interlayer hopping $t_\perp$.
(b)~Projector-weighted layer contributions ($\sigma^{yxx}_{\rm t}$, $\sigma^{yxx}_{\rm b}$) and total
($\sigma^{yxx}_{\rm tot}$) shift current versus chemical potential $\mu$ at
$\Delta_E=0$. (c)~Raw total shift current $\sigma^{yxx}_{\rm tot}$ versus $\mu$ at
$+\Delta_E$ and $-\Delta_E$. (d)~Raw total shift current at fixed chemical potential
$\mu_\ast$ versus the layer-staggered potential $\Delta_E$. Parameters:
$m=1.20$, $\beta=0.40$, $\gamma=0.60$, $\Delta_0=0.45$, $t_\perp=0.03$,
$|\Delta_E|=0.26$~eV, with $\mu_\ast\simeq0.53$~eV.}
\end{figure}

\textit{Minimal 2D model}---We illustrate the shift-current LTE 
($V[V^2]$) with a minimal bilayer tight-binding model. The top layer is a spinless two-band 
system that preserves the mirror operation $M_x:(k_x,k_y)\to(-k_x,k_y)$. We take
\begin{equation}
  H_{\rm t}(\mathbf{k})
  =d_x(\mathbf{k})\rho_x+d_y(\mathbf{k})\rho_y
   +d_z(\mathbf{k})\rho_z ,
\label{eq:2d-top}
\end{equation}
with
\begin{equation}
\begin{split}
  d_x&=m+\cos k_x+\cos k_y,\\
  d_y&=\sin k_y,\\
  d_z&=\Delta_0+\beta\cos k_x+\gamma\sin k_y .
\end{split}
\label{eq:2d-dvector}
\end{equation}
Here $\rho_i$ are Pauli matrices in the orbital subspace. The term
$\gamma\sin k_y$ breaks $M_y$ while preserving $M_x$, thereby producing a
$y$-polar distortion in the top layer. Under the $C_{2x}$ layer-exchange
operation, $k_y\to-k_y$, this term changes sign. The bottom layer therefore
carries the opposite polar distortion ($\pm P_y$), as illustrated in Fig.~\ref{fig:2d-model}(a), with its Hamiltonian given by
\begin{equation}
H_{\rm b}(k_x,k_y)
=
H_{\rm t}(k_x,-k_y) .
\label{eq:2d-bottom}
\end{equation}
Including a symmetry-preserving interlayer hopping gives the bilayer Hamiltonian:
\begin{equation}
  H^0_{\rm bi}(\mathbf{k})=
  \begin{pmatrix}
    H_{\rm t}(\mathbf{k}) & t_\perp I_2\\
    t_\perp I_2 & H_{\rm b}(\mathbf{k})
  \end{pmatrix},
\label{eq:2d-bilayer}
\end{equation}
where $I_2$ is the identity matrix in orbital space and $t_\perp$ is an
orbital-diagonal interlayer hopping. We represent the effect of an
out-of-plane electric field by an effective layer-staggered potential
$\Delta_E$, which shifts the two layers by $\pm\Delta_E$,
\begin{equation}
  H_{\rm bi}(\mathbf{k},\Delta_E)=H^0_{\rm bi}(\mathbf{k})
  +\Delta_E\,\tau_z\otimes I_2 ,
\label{eq:2d-field}
\end{equation}
where $\tau_i$ are Pauli matrices in layer space.  The layer-exchange operator
$\mathcal{L}_x=\tau_x\otimes I_2$ then satisfies
\begin{equation}
\mathcal{L}_x
H_{\rm bi}(k_x,k_y,\Delta_E)
\mathcal{L}_x^{-1}
=
H_{\rm bi}(k_x,-k_y,-\Delta_E).
\label{eq:layer-exchange-model}
\end{equation}
Thus, the intrinsic bilayer has an exact layer-exchange symmetry.

Here we consider the shift current component $\sigma^{yxx}$.
Under $C_{2x}$, the current index $y$ changes sign, while the two $x$-polarized optical-field indices are unchanged. The two projected layer responses therefore obey
\begin{equation}
\sigma^{yxx}_{\rm b}(\mu,0)
=
-\sigma^{yxx}_{\rm t}(\mu,0),
\label{eq:layer-cancel-model}
\end{equation}
so that the total bilayer response vanishes at zero field. At finite
$\Delta_E$,
Eq.~\eqref{eq:layer-exchange-model} instead relates opposite field polarities and enforces
\begin{equation}
\sigma^{yxx}_{\rm bi}(\mu,\Delta_E)
=
-\sigma^{yxx}_{\rm bi}(\mu,-\Delta_E).
\label{eq:odd-field-model}
\end{equation}
We evaluate the independent-particle shift-current tensor
\cite{Sipe2000,Morimoto2016Shift,Cook2017Shift}
for the hybridized four-band Hamiltonian and retain the directly calculated total response while using a projector-weighted layer decomposition for visualization. The formula and projection procedure are given in Sec.~IV of the SM~\cite{SM}. The independently calculated raw responses in Fig.~\ref{fig:2d-model} verify these symmetry constraints. As shown in Fig.~\ref{fig:2d-model}(b), the two projected layer
contributions cancel to numerical precision at $\Delta_E=0$, yielding $\sigma^{yxx}_{\rm bi}(\mu,0)=0$ despite finite interlayer hybridization. Figures~\ref{fig:2d-model}(c) and~\ref{fig:2d-model}(d) further show that the field-induced response
is antisymmetric under $\Delta_E\to -\Delta_E$ over the plotted ranges of chemical potential and layer-staggered potential. The model therefore realizes a hidden layer shift-current channel whose sign is electrically switchable by reversing the out-of-plane field. The tensor component $\sigma^{yxx}$ is even under $M_x$ but odd under $M_y$. Therefore a finite $\sigma^{yxx}$ requires $M_y$ to be broken while $M_x$ is preserved.

\begin{figure}[!tbp]
  \centering
  \includegraphics[width=\columnwidth]{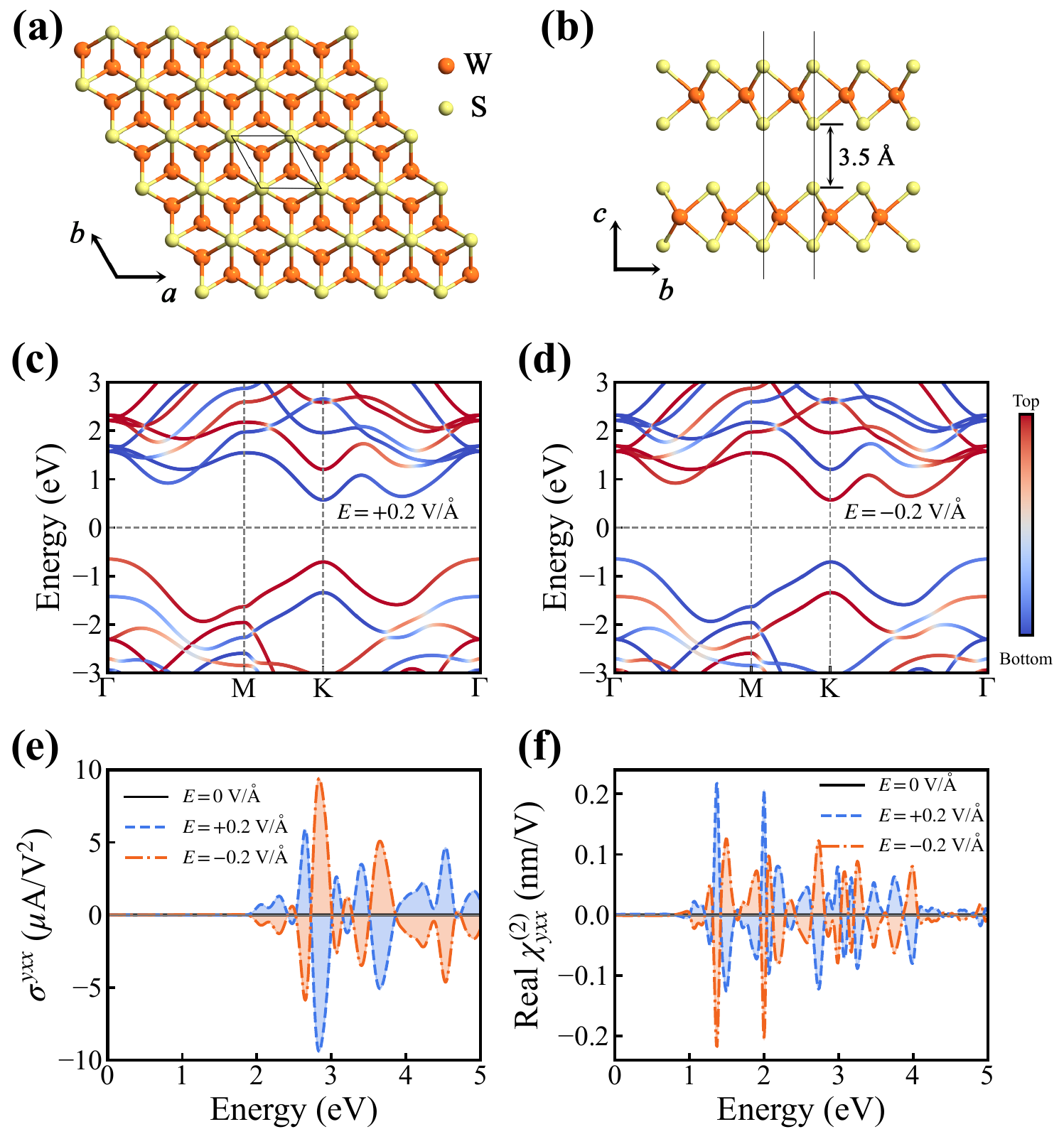}
  \caption{(a) Top and (b) side views of the crystal structure of bilayer $\text{WS}_2$ related by spatial inversion $\mathcal{P}$. Calculated layer-resolved band structures under an applied out-of-plane electric field of (c) $E_z = +0.2$ V/\AA{} and (d) $E_z = -0.2$ V/\AA{}. Photon-energy dependence of (e) the shift current tensor component $\sigma^{yxx}$ and (f) the second-harmonic generation tensor component $\text{Re}[\chi^{(2)}_{yxx}]$ at zero electric field and out-of-plane electric fields of $E_z = \pm 0.2$ V/\AA{}.}
  \label{fig:WS2}
\end{figure}

\textit{Material realization}---Guided by the symmetry conditions outlined in Tab.~\ref{tab:stacking}, we propose that the inversion-symmetric ($\mathcal{P}$) stacked bilayer $\text{WS}_2$ serves as a material candidate exhibiting electrically switchable layer shift current and second-harmonic generation (SHG) tensors~\cite{Klein2017FieldSHG,Jiang2022InterlayerSHG}. Monolayer $\text{WS}_2$ belongs to the crystallographic layer group $p\bar{6}m2$ (No.~78) with optimized lattice constants $a = b = 3.17~\text{\AA}$. As illustrated in Figs.~\ref{fig:WS2}(a) and~\ref{fig:WS2}(b), the bilayer $\text{WS}_2$ related by spatial inversion $\mathcal{P}$ is centrosymmetric, crystallizing in the layer group $p\bar{3}m1$ (No. 72). 
Owing to $\mathcal{PT}$ symmetry, the electronic states derived from the two layers are degenerate at the whole Brillouin zone. By applying an out-of-plane electric field, the symmetry can be broken and the degeneracy is lifted.
Figures~\ref{fig:WS2}(c) and~\ref{fig:WS2}(d) display the layer-projected band structures of bilayer $\text{WS}_2$ under out-of-plane electric fields of $\pm E_z$ obtained from first-principles calculations (see Sec.~V of the SM~\cite{SM} for computational details). The applied electric field reduces the layer group of the bilayer to $p3m1$ (No.~69).

Subsequently, we evaluate the shift current and SHG of bilayer $\text{WS}_2$ under positive and negative electric fields. Note that for the $C_{3v}$ point group governing the bilayer under an electric field, the purely in-plane components of both the shift current and SHG tensors possess only a single independent component, satisfying $\sigma^{yxx} = \sigma^{xyx} = \sigma^{xxy} = -\sigma^{yyy}$ and $\chi^{(2)}_{yxx} = \chi^{(2)}_{xyx} = \chi^{(2)}_{xxy} = -\chi^{(2)}_{yyy}$. As shown in Figs.~\ref{fig:WS2}(e) and~\ref{fig:WS2}(f), in the absence of an external electric field, inversion symmetry strictly enforces the macroscopic shift current ($\sigma^{yxx}$) and SHG ($\chi^{(2)}_{yxx}$) of bilayer $\text{WS}_2$ to vanish. In stark contrast, the out-of-plane electric field yields finite $\sigma^{yxx}$ and $\chi^{(2)}_{yxx}$ signals. Furthermore, the responses under $+E_z$ and $-E_z$ are equal in magnitude but opposite in sign. This field-reversal of the shift current and SHG confirms the predicted electrically switchable layer tensor response, in agreement with the symmetry analysis. Both demonstrations employ a \emph{unitary} layer-exchange operation ($C_{2x}$ for the minimal model and $\mathcal{P}$ for bilayer $\text{WS}_2$); the resulting field-odd responses follow from the spatial selection rule alone and are therefore independent of the Onsager considerations in Sec.~II of the SM~\cite{SM}.

To emphasize the experimentally measurable effects for LTE, here we use three representative responses and corresponding tensors to illustrate how these predictions can be connected to existing experimental techniques. First, shift-current tensors have recently been extracted from polarization-resolved photocurrent spectra in multiterminal monolayer $\mathrm{MoS}_2$ devices, with substantial electrostatic tunability \cite{Chen2026ExcitonicShift}. For bilayer $\mathrm{WS}_2$, we propose an encapsulated dual-gate device in which opposite perpendicular displacement fields are applied at fixed carrier density. Measuring the signed photocurrent under unchanged linearly polarized illumination and zero source--drain bias would test the predicted reversal of $\sigma^{yxx}$. Its polarization dependence and spatial distribution would help distinguish the intrinsic response from contact and photothermal contributions.

Second, electric-field-induced SHG has already been demonstrated in centrosymmetric bilayer $\mathrm{MoS}_2$ \cite{Klein2017FieldSHG}. Dual-gated bilayer $\mathrm{WSe}_2$ experiments have also independently controlled carrier density and perpendicular field, enhancing SHG through a layer-dependent exciton-polaron effect \cite{Cha2024ExcitonPolaron}. These studies provide realistic device platforms for testing our $\mathrm{WS}_2$ prediction. Polarization-resolved measurements can identify the tensor components, while interferometric detection can resolve the $\pi$ phase reversal associated with $\chi^{(2)}(+E_z)=-\chi^{(2)}(-E_z)$. Such phase-sensitive detection has been demonstrated in two-dimensional TMDs \cite{Zhao2026InterferometricSHG}. Conventional intensity measurements alone cannot resolve this overall sign reversal.

Third, the experimentally observed odd--even layer dependence of piezoelectricity in $\mathrm{MoS}_2$ provides an electromechanical example of interlayer cancellation \cite{Wu2014Piezo,Zhu2015Piezo}. A possible test of the piezoelectric LTE is to apply uniform oscillatory in-plane strain to a centrosymmetric bilayer under opposite perpendicular fields and measure the strain-induced charge at lateral electrodes. A field-reversed piezoelectric coefficient would produce a reversal of the strain-synchronized signal. These experimental precedents provide practical routes for detecting LTE channels. While our symmetry analysis establishes the field-reversal relations, the magnitude and observability of the responses depend on material properties and experimental conditions. Experimental verification therefore requires separating the intrinsic LTE response from background contributions associated with contacts, substrates, and other device-dependent factors.

\textit{Discussion}---We have established the LTE as a general symmetry principle that extends the layer Hall effect to a broad class of tensor responses in stacked systems. Once the local response is classified by its Jahn symbol, the action of a layer-exchanging operation determines the layer parity of each tensor component. A layer-odd component can remain finite locally while canceling between symmetry-related layers, thereby forming a hidden tensor response. An out-of-plane electric field breaks the equivalence of the two layers and allows a macroscopic response whose sign reverses with the field. This construction places optical, electromechanical, magnetic, and transport responses within the same symmetry framework and separates the universal selection rules from the material-dependent magnitude and spectral structure of the response.

In this paper, we find that electrical reversibility is determined by the transformation of the response tensor under the layer-exchange operation. This establishes which components cancel at zero field and which symmetry-allowed components reverse under opposite fields. The resulting switching relations remain valid with finite interlayer hybridization, although the response magnitudes depend on microscopic details. This principle enables response-selective electrical control for those tensors that were previously difficult to manipulate. For example, in an inversion-related stack, polar rank-three tensors describing shift current and electric-dipole SHG are field-odd, whereas the rank-two longitudinal conductivity is field-even. Thus, comparing equal and opposite perpendicular fields at fixed carrier density can reverse the nonlinear optical responses while yielding equal longitudinal conductivity. This illustrates how different functionalities can be controlled selectively within the same structure.

Our first-principles results for bilayer $\mathrm{WS}_2$ provide a concrete realization of these optical switching relations. Reversing the perpendicular field switches the shift-current direction and the sign of the SHG susceptibility; the latter corresponds to a phase reversal of the generated second-harmonic field. These responses provide complementary electrical and optical readout channels for the electrically selected layer configuration. The framework therefore supplies a symmetry-based design rule for choosing stackings and response components with the desired switching behavior, together with specific material predictions demonstrating the resulting functionality.

We emphasize that the proposed layer-parity criterion does not rely on a strict decoupled-layer approximation. Finite interlayer hybridization is fully permissible within this symmetry framework: it may redistribute the response weight across individual layers and modify the overall magnitude of the effect, but it cannot lift the symmetry-enforced cancellation of a layer-odd response channel so long as the underlying layer-exchange symmetry remains unbroken. This robust combination of symmetry-protected qualitative selection rules and quantitatively tunable responses via interlayer coupling opens promising avenues for electrically and mechanically reconfigurable van der Waals devices, where a single stacking degree of freedom can simultaneously modulate a wide range of optical~\cite{Zheng2023BilayerShift,Atlam2025BilayerShift,lte_dong_layer_pv_2026,layer16,Penaranda2024TBGPhoto,Chaudhary2022TBGShift}, transport~\cite{lte_wang_qmetric_2023,lte_gao_qmetric_2023,Zhao2023BCP,Zhao2024NCT}, magnetic~\cite{lte_dong_layer_pv_2026,Bennett2024StackingFE,Wang2020StackingDomainMagnons}, and electromechanical~\cite{lte_wan_res2_2022,Zhou2024StackingFE,Bennett2024StackingFE} responses. Layer-selective contacts~\cite{Wang2025LayerSelectiveContact,Kegelmann2017DoubleLayerContacts}, layer-selective electronic and chemical engineering~\cite{Seiler2025LayerSelectiveSOC,Tong2026LayerSelectiveHydrogenation,Zhong2023LayerSelectiveQSH}, dual-gate geometries~\cite{experiment6,experiment10,NLHE18,Chen2016DualGate}, and stacking-domain engineering~\cite{Wang2020StackingDomainMagnons,Jiang2018DomainWallSolitons} can further be exploited to activate, filter, and read out these otherwise hidden tensor channels, offering a symmetry-guided design strategy for multifunctional van der Waals devices.

The framework can be extended beyond ordinary magnetic point groups to the spin-group description recently developed for nonrelativistic collinear magnets, including altermagnets~\cite{work1,He2023TwistedBilayerSpin,Pan2025FloquetAltermagnet,Etxebarria2025,Elcoro2026}. In this setting, spatial operations may be combined with independent spin rotations, so that the spatial tensor character, spin character, time-reversal parity, and layer parity of a response can be classified on equal footing~\cite{Huang2026SpinFerroelasticity}. Such an extension would enable the systematic identification of hidden layer-resolved spin currents, spin polarizations, and spin-dependent optical responses that are not fully captured by magnetic point groups alone~\cite{LHE5}. In moiré and sliding-ferroelectric systems, relative translation or twist can further modify the layer-exchange coset itself, allowing structural motion not only to reverse a response but also to switch between distinct symmetry-allowed tensor components~\cite{layer4,Xiao2022SlidingBPVE,Wang2024UltrafastSlidingPolarization}.

The LTE framework can be extended to multilayers by identifying a symmetry operation of the entire stack that connects its configurations under opposite perpendicular electric fields. The multilayer stacking theory of Xin, Guo, et al \cite{Xin2026Multilayer} provides a constructive approach to finding this operation. Their construction adds two outer layers to a central multilayer block. Every surviving symmetry operation must preserve the central block and be compatible with the added layers and their translation operation. Operations that preserve the out-of-plane coordinate leave each outer layer unchanged, whereas operations that reverse it exchange the outer layers. This procedure applies recursively along separate odd- and even-layer sequences starting from a monolayer and a bilayer, respectively.

For LTE, the key step is to identify a surviving operation that reverses the perpendicular field and maps the entire stack between opposite-field configurations. A tensor component or linear combination that changes sign under this operation vanishes at zero field and reverses under opposite fields. A finite response must additionally be allowed by the symmetry remaining under bias. These conclusions apply independently of layer number and remain valid with finite interlayer hybridization. Multilayer LTE design therefore requires identifying the operation connecting opposite-field states and testing the transformation of the selected tensor; layer number or structural polarity alone does not determine electrical reversibility.

\begin{acknowledgments}
This work is supported by the National Natural Science Foundation of China (Grant Nos.~12574070 and 12504223), the China Postdoctoral Science Foundation (2025M773383 and GZC20252231), the China Postdoctoral
Science Foundation-Hunan Joint Support Program (2025T002HN), the Scientific Research Fund of Hunan Provincial Education Department (24B0658), and the Science Foundation of Hengyang Normal University of China (Grant No. 2024QD38).
\end{acknowledgments}

\indent Zehou Li, Shenda He and Pan Zhou contributed equally to this work.
\nocite{Aversa1995,IbanezAzpiroz2018,abacus1,abacus2,GGA,DFT-D3,NAO,SG15,ONCV,PYATB}
\bibliography{references}

\end{document}